\documentclass[preprint,superscriptaddress,showkeys,showpacs,floatfix,eqsecnum,onecolumn]{revtex4}
\usepackage{graphics}
\usepackage{nicefrac}
\usepackage{graphicx}
\usepackage{dcolumn}
\usepackage{bm}
\usepackage{amssymb,amsmath,graphicx,graphics,color,epstopdf,mathtools}
\usepackage{caption}
\usepackage{hyperref}
\usepackage[capitalise]{cleveref}
\usepackage{rotfloat}

\hypersetup{
    colorlinks=true,
    linkcolor=blue,
    filecolor=magenta, citecolor=blue,    urlcolor=blue,
}

\begin{document}
\preprint{HEP/123-qed}
\title{Greybody Radiation and Quasinormal Modes of Kerr-like Black Hole in Bumblebee Gravity Model}
\author{Sara Kanzi}
\affiliation{}
\author{}
\affiliation{}
\author{\.{I}zzet Sakall{\i}}
\affiliation{Physics Department, Eastern Mediterranean
University, Famagusta, North Cyprus via Mersin 10, Turkey.}
\author{}
\affiliation{}

\begin{abstract}
In the framework of the Lorentz symmetry breaking (LSB), we investigate the quasinormal modes (QNMs) and the greybody factors (GFs) of the Kerr-like black hole spacetime obtained from the bumblebee gravity model. In particular, we analyze the scalar and fermionic perturbations of the black hole within the framework of both semi-analytic WKB method and the time domain approach. The impacts of the LSB on the bosonic/fermionic QNMs and GFs of the Kerr-like black hole are investigated in detail. The obtained results are graphically depicted and discussed.

\end{abstract}
\volumeyear{ }
\eid{ }
\date{\today}
\received{}

\maketitle
\tableofcontents

\section{Introduction}

One of the cornerstones of modern physics is the Lorentz invariance, which is a fundamental part of both general relativity (GR) and the standard model of particle physics. However, today there are several theories that Lorentz invariance can not be valid
at all energies \cite{1}. Lorentz invariance violation (LIV) may yield a glimpse of quantum gravity (QG). Although from the theoretical point of view the exploration of this possibility has been active for many years [see for instance \cite{IS1}, and references therein], the phenomenology of LIV has been developed only within the last decade \cite{2,3,4}. Before the mid-1990s, there were only few works about the experimental consequences of LIV \cite{IS2,IS3,IS4}, because the new effects were expected only in the particle interactions, which occur at energies of Planck mass: $M_{\mathrm{Pl}} \equiv \sqrt{\hbar c / G_{N}} \simeq 1.22 \times 10^{19} \mathrm{GeV} / c^{2}$. Then, it was realized that there exists special cases in which the new effects could appear also at lower energies. Those particular cases have opened "windows on the QG". Nowadays, this subject has been investigating in various fields of QG: string theory \cite{5}, loop QG theory \cite{6}, and even in the non-gravity theory \cite{7}.

Standard model extension (SEM ) \cite{8,9,10} is an effective field theory which describing the standard model coupled to GR with the Lorentz and CPT invariance violations. It is worth noting that CPT invariance requires the physics to be unchanged under the combination of charge conjugation (C), parity inversion (P), and time reversal (T). In SEM, other important consequences can arise, like the
appearance of Nambu-Goldstone and Higgs modes. Unlike the effective framework
provided by the SME, the properties of these modes are, in general, model dependent and
cannot be completely discussed without knowledge of the underlying fundamental theory. On the other hand, the most studied LIV models that contemplate the role of the extra modes arising from the LSB mechanism involve the vacuum condensation of a vector field. These models are called “bumblebee models” \cite{11,12,13}, which were first introduced by Kostelecky and Samuel in 1989 \cite{14,s14}. This model was then extended to the various fields including the gravity \cite{s15,k15,l15}. Remarkably, in 2018, an exact Schwarzschild-like solution in this bumblebee gravity model (BGM) was found by Casana et al \cite{l14}. Following this study, the accretion onto that black hole was studied by Rong-Jia Yang et al \cite{d14} who showed that the LSB parameter slows down the mass
accretion rate. Moreover, the problems of gravitational lensing and Hawking radiation of the black hole have been recently studied in \cite{16} and \cite{17}, respectively. On the other hand, it is a fact accepted by researchers that rotating black hole spacetimes are
the most relevant sub-cases for astrophysics. Such solutions might also describe exterior metric for the rotating stars. The Kerr-like black solution in the BGM has been recently obtained by Ding et al \cite{18}. This stationary, axisymmetric, and asymptotically flat $4$-dimensional Kerr-like black hole solution prescribed in the bumblebee gravity theory is obtained with a bumblebee vector field, which is coupled to the spacetime curvature and acquires a vacuum expectation value that induces the LSB. When the rotation $a \rightarrow 0$, then one can recover the Schwarzschild-like black solution \cite{l14}. Furthermore, if LIV or LSB constant vanishes, $\ell \rightarrow 0,$ then the well-known Kerr black hole solution is recovered. Since a perturbed black hole emits gravitational radiation from its
horizon, which reveals information about its inner properties \cite{19}, one can study the quantum structure of the black holes. Outside the horizon, potential barrier acts as a filter which depends on the frequencies of the propagating waves. Some waves
are reflected by the barrier and some transmitted to infinity \cite{20}. However, the
observers at infinity receives only a fraction of the emitted radiation. Therefore, there is a deviation between the radiation emanating from the black hole's horizon and the observed radiation. This phenomenon also manifests itself in the Hawking radiation \cite{H21,H22}. Namely, Hawking radiation which is modified from the perfect black body spectrum is known as the GF \cite{21,22}. There are
different methods to compute the GF such as the WKB approximation \cite{23,24}, matching method \cite{25,26}, rigorous bound method \cite{27}, and analytical method for the various of spin fields \cite{28,y28,d28,h28,g28,l28}. Furthermore, in the framework of GR for the radiation of
gravitational waves, the most important phase described in function of the
proper oscillation frequencies of the black hole is called QNMs, which depend on the black hole parameters \cite{d29,u29}. The propagation of QNMs is different from normal modes. Namely, they possess a unique complex frequency spectrum in which its real part represents the frequency of the oscillation and the imaginary part shows the damping \cite{f29,s29}. One of the appropriate methods to compute the QNM is the WKB approximation
\cite{Sara34,Sara35}, which is a
semi-analytic technique. This method was also studied within the different contexts such as the AdS/CFT correspondence \cite{r29,t29,tr29}, the black hole spectroscopy, the black hole quantization
\cite{g29,w29,x29}, and quantum singularity of black holes, in particular for the rotating ones such as the Kerr black hole \cite{z29,j29,b29}.

This paper is organized as follows: In Sec. \eqref{sec2}, we review the
Kerr-like spacetime of the BGM. Sections \eqref{sec3} and \eqref{sec4} are devoted to the scalar and fermionic perturbations in the Kerr-like black hole geometry, respectively. To this end , we first derive the associated effective potentials of bosons and
fermions by using the Klein-Gordon and Dirac equations, respectively. Then, we study the GFs of both particles from the black hole in Sec. \eqref{sec5}. Bosonic and fermionic QNMs of the Kerr-like black hole are numerically computed with the aid of the WKB method in Sec. \eqref{sec6}. Finally, we draw our conclusions in Sec. \eqref{sec7}. 

(\textit{Throughout the paper, we follow the metric convention $(-,+,+,+)$ and the geometrized units: $G = c = \hbar = 1$.})
\newpage

\section{Kerr-like Black Hole Spacetime of BGM} \label{sec2}

In the bumblebee gravity theory, action of an electromagnetic field coupled
to the bumblebee vector field for the curved spacetime is given by \cite{18}%

\begin{equation}
S=\int d^{4}x\sqrt{-g}\left[  \frac{1}{16\pi G_{N}}\left(  \Re+\varrho B^{\mu
}B^{\upsilon}\Re_{\mu\nu}\right)  -\frac{1}{4}B^{\mu\nu}B_{\mu\nu}-V\left(
B^{\mu}\right)  \right]  , \label{sk1}%
\end{equation}
where $\varrho$ is coupling constant and $B_{\mu}$ is bumblebee field with
mass dimension-$1$, which requires a non-zero vacuum expectation value as
$\langle B^{\mu}\rangle=b^{\mu}$. The bumblebee field strength is defined as follows%
\begin{equation}
B_{\mu\upsilon}=\partial_{\mu}B_{\nu}-\partial_{\nu}B_{\mu}. \label{sk2}%
\end{equation}

The function of potential $V\left(  B^{\mu}\right)  $ is given by \cite{10,s28}%

\begin{equation}
V=V\left(  B_{\mu}B^{\mu}\pm b^{2}\right)  , \label{sk3}%
\end{equation}
where $b$ is a real constant. The Kerr-like rotating black hole metric in the BGM was recently found by \cite{18} as follows%
\begin{equation}
ds^{2}=-\left(  1-\frac{2Mr}{\rho^{2}}\right)  dt^{2}-\frac{4Mra\sqrt{1+L}%
\sin^{2}\left(  \theta\right)  }{\rho^{2}}dtd\phi+\frac{\rho^{2}}{\Delta
}dr^{2}+\rho^{2}d\theta^{2}+\frac{A\sin^{2}\left(  \theta\right)  }{\rho^{2}%
}d\varphi^{2},\label{1}%
\end{equation}
where%

\begin{equation}
\Delta=\frac{r^{2}-2Mr}{1+L}+a^{2},\quad \rho^{2}=r^{2}+\left(  1+L\right)
a^{2}\cos^{2}\left(  \theta\right)  , \label{2}%
\end{equation}

\begin{equation}
A=\left[  r^{2}+\left(  1+L\right)  a^{2}\right]  ^{2}-\Delta\left(
1+L\right)  ^{2}a^{2}\sin^{2}\left(  \theta\right)  .\label{3}%
\end{equation}

One can immediately see that as $L\rightarrow0$ in the metric (\ref{1}), the spacetime reduces to the metric of well-known Kerr black hole. Besides, when $a\rightarrow0$ it represents the Schwarzschild-like solution having the LSB \cite{l14}
\begin{equation}
ds^{2}=-\left(  1-\frac{2M}{r}\right)  dt^{2}+\frac{1+L}{1-2M/r}dr^{2}%
+r^{2}d\theta^{2}+r^{2}\sin^{2}\theta.\label{4}%
\end{equation}

In short, metric (\ref{1}) is nothing but a solution of LIV black hole with a rotation parameter, which is equal to the angular momentum per unit mass: $a=\frac{J}{M}$. Its singularities appear at $\rho^{2}=0$ and $\Delta=0$. For $\rho^{2}=0$, we have a ring-shape physical singularity at the equatorial plane of the center of rotating black hole having radius $a$. The roots of Eq. (\ref{2}) reveal the locations of the event horizon and ergosphere:
\begin{equation}
r_{\pm}=M\pm\sqrt{M^{2}-a^{2}\left(  1+L\right)  },\text{ \ \ \ \ \ \ \ \ \ }%
r_{\pm}^{ergo}=M\pm\sqrt{M^{2}-a^{2}\left(  1+L\right)  \cos^{2}\theta
}, \label{5}%
\end{equation}
in which $\pm$\ signs indicate the outer and inner horizon/ergosphere,
respectively. For having a black hole solution, it is conditional on 
\begin{equation}
a\leq\frac{M}{\sqrt{1+L}}.\label{6}%
\end{equation}

Now, we can write the metric tensor of the Kerr-like spacetime as%

\begin{equation}
g_{\mu\upsilon}=\left(
\begin{array}
[c]{cccc}%
-\left(  1-\frac{2Mr}{\rho^{2}}\right)  & 0 & 0 & -\frac{2Mra\sqrt{1+L}%
\sin^{2}\theta}{\rho^{2}}\\
0 & \frac{\rho^{2}}{\Delta} & 0 & 0\\
0 & 0 & \rho^{2} & 0\\
-\frac{2Mra\sqrt{1+L}\sin^{2}\theta}{\rho^{2}} & 0 & 0 & \frac{A\sin^{2}%
\theta}{\rho^{2}}%
\end{array}
\right),  \label{7}%
\end{equation}
from which one can compute the determinant of the metric tensor as follows
\begin{equation}
g\equiv\det\left(  g_{\mu\nu}\right)  =-\rho^{4}\left(  1+L\right)  \sin
^{2}\theta. \label{8}%
\end{equation}

Thus, the contravariant form of $g_{\mu\nu}$ can be easily obtained as%
\begin{equation}
g^{\mu\nu}=\left(
\begin{array}
[c]{cccc}%
-\frac{A}{\rho^{2}\Delta\left(  1+L\right)  } & 0 & 0 & -\frac{2Mra}{\rho
^{2}\Delta\sqrt{1+L}}\\
0 & \frac{\Delta}{\rho^{2}} & 0 & 0\\
0 & 0 & \frac{1}{\rho^{2}} & 0\\
-\frac{2Mra}{\rho^{2}\Delta\sqrt{1+L}} & 0 & 0 & \frac{\rho^{2}-2Mr}{\rho
^{2}\Delta\left(  1+L\right)  \sin^{2}\theta}%
\end{array}
\right).  \label{9}%
\end{equation}

One can also get the Hawking temperature of this Kerr-like black hole which is derived from its surface gravity ($\kappa$) \cite{18} as follows
\begin{eqnarray}
T=\frac{\kappa}{2 \pi}, \medspace \kappa=-\frac{1}{2} \lim _{r \rightarrow
r_{+}} \sqrt{\frac{-1}{Y}} \frac{dY}{d r}, \medspace Y \equiv g_{t
t}-\frac{g_{t \varphi}^{2}}{g_{\varphi \varphi}}. \label{8new1}
\end{eqnarray}

By using the relevant metric components given in Eq. (\ref{7}) and substituting them into Eq. (\ref{8new1}), the Hawking temperature is found to be
\begin{eqnarray}
T=\frac{\sqrt{M^2-a^2\left(1+L\right)}} {4\pi M
\sqrt{1+L}\left(M+\sqrt{M^2
-a^2\left(1+L\right)}\right)}. \label{8new2}
\end{eqnarray}

\section{Scalar Perturbations} \label{sec3}

In this section, we shall examine the scalar perturbations of the Kerr-like black hole and derive the effective potential to which scalar waves will be exposed in this geometry. To this end, we employ the Klein-Gordon equation:
\begin{equation}
\frac{1}{\sqrt{-g}}\partial_{\mu}\left(  \sqrt{-g}g^{\mu\nu}\partial_{\nu
}\right)  \Psi=\mu_{0}^{2}\Psi,\label{10}%
\end{equation}
where $\mu_{0}$ is the mass of the scalar particle. Using Eqs. (\ref{8}) and (\ref{9}) in Eq. (\ref{10}), we get%
\begin{multline}
-\frac{A}{\rho^{2}\Delta\left(  1+L\right)  }\partial_{t}^{2}\Psi-\frac
{2Mra}{\Delta\rho^{2}\sqrt{1+L}}\partial_{t}\partial_{\phi}\Psi+\frac{1}%
{\rho^{2}\sin\theta}\partial_{\theta}\left(  \sin\theta\partial_{\theta
}\right)  \Psi+\\
\frac{1}{\rho^{2}}\partial_{r}\left(  \Delta\partial_{r}\right)  -\frac
{2Mra}{\rho^{2}\Delta\sqrt{1+L}}\partial_{\phi}\partial_{t}\Psi+\frac{\rho
^{2}-2Mr}{\Delta\rho^{2}\left(  1+L\right)  \sin^{2}\theta}\partial_{\phi}%
^{2}\Psi=\mu_{0}^{2}\Psi\label{11}.
\end{multline}

To apply the technique of separation of variables in Eq. (\ref{11}), one can
use the following ansatz:
\begin{equation}
\Psi\left(  r,t\right)  =R\left(  r\right)  S\left(  \theta\right)  e^{im\phi
}e^{-i\omega t}, \label{12}%
\end{equation}
where $m$ is azimuthal quantum number and $\omega$ represents the
energy of the particles. Therefore, the radial equation becomes
\begin{multline}
\frac{1}{R\left(  r\right)  }\frac{d}{dr}\left(  \Delta\frac{dR\left(
r\right)  }{dr}\right)  +\frac{\omega^{2}\left(  r^{2}+\left(  1+L\right)
a^{2}\right)  ^{2}}{\Delta\left(  1+L\right)  }+\\
\frac{m^{2}a^{2}}{\Delta}-\frac{4Mram\omega}{\Delta\sqrt{1+L}}-\omega^{2}%
a^{2}\left(  1+L\right)  -\mu_{0}^{2}r^{2},\label{S13}
\end{multline}
and the angular part reads%
\begin{equation}
\frac{1}{S\left(  \theta\right)  \sin\theta}\frac{d}{d\theta}\left(
\sin\theta\frac{dS\left(  \theta\right)  }{d\theta}\right)  -\frac{m^{2}}%
{\sin^{2}\theta}+c^{2}\cos^{2}\theta. \label{14}%
\end{equation}

As is known, angular and radial equations admit two same (in absolute) eigenvalues with opposite signs. Angular differential equation (\ref{14}) has solutions in terms of the oblate spherical harmonic functions
$S_{lm}\left(  ic,\cos\theta\right)  $ having eigenvalue $\lambda_{lm}$ \cite{Sara30} in which
$l,m$ are integers such that $\left\vert m\right\vert \leq l,$ and $c^{2}%
=a^{2}\left(  1+L\right)  \left(  \omega^{2}-\mu_{0}^{2}\right)  $ \cite{Sara31}. For simplicity, we consider the separation constant as $\lambda=\lambda_{lm}$. Thus, the radial differential equation becomes
\begin{multline}
\Delta\frac{d}{dr}\left(  \Delta\frac{dR\left(  r\right)  }{dr}\right)
+\left\{  m^{2}a^{2}+\frac{\omega^{2}}{\left(  1+L\right)  }\left(
r^{2}+\left(  1+L\right)  a^{2}\right)  ^{2}-\right.  \label{S15}\\
\left.  \frac{4Mram\omega}{\sqrt{1+L}}-\left(  \mu_{0}^{2}r^{2}+\omega
^{2}a^{2}\left(  1+L\right)  +\lambda\right)  \Delta\right\}  R\left(
r\right)=0.
\end{multline}

The radial solution is in general associated with a free oscillation mode of the propagating field. Stable modes have particular frequencies $\omega$ with complex negative imaginary values: so we have an exponentially subsidence in amplitude. Conversely, if the imaginary values of the frequencies are positive, then the amplitude of the oscillations exponentially increase and the modes consequently become unstable. If we consider $M\omega\ll1$ and $\mu M\ll1$, which was first noticed by Starobinskii \cite{Sara32,Sara31}, then Eq. (\ref{14}) is amenable to analytic methods. If we assume the inequalities to hold then the angular part can be thought as spherical harmonics with $\lambda\simeq l\left(  l+1\right)$.

For having one dimensional wave equation, we first use the following transformation%
\begin{equation}
R\left(  r\right)  =\frac{U\left(  r\right)  }{\sqrt{r^{2}+\left(  1+L\right)
a^{2}}},\label{16}%
\end{equation}
together with the tortoise coordinate:
\begin{equation}
\frac{dr_{\ast}}{dr}=\frac{r^{2}+\left(  1+L\right)  a^{2}}{\sqrt{1+L}\Delta
}. \label{17}%
\end{equation}

Thus, Eq. (\ref{S15}) can be expressed as a one-dimensional Schr\"{o}dinger equation 

\begin{equation}
\frac{d^{2}U}{dr_{\ast}^{2}}+\left(  \omega^{2}-V_{eff}\right)  U=0, \label{19}%
\end{equation}
where the effective potential reads%
\begin{multline}
V_{eff}=\frac{\left(  1+L\right)  \Delta}{\left(  r^{2}+\left(  1+L\right)
a^{2}\right)  ^{2}}\times\label{20}\\
\left[  \frac{\Delta\acute{}r}{\left(  r^{2}+\left(  1+L\right)  a^{2}\right)  }+\frac{\Delta}{\left(
r^{2}+\left(  1+L\right)  a^{2}\right)  }-\frac{3r^{2}\Delta}{\left(
r^{2}+\left(  1+L\right)  a^{2}\right)  ^{2}}\right.  \\
\left.  +\frac{4Mram\omega}{\Delta\sqrt{1+L}}-\frac{m^{2}a^{2}}{\Delta
}+\left(  \mu_{0}^{2}r^{2}+\omega^{2}a^{2}\left(  1+L\right)  +\lambda\right)
\right].
\end{multline}

\begin{figure}[h]
\centering\includegraphics[scale=0.6]{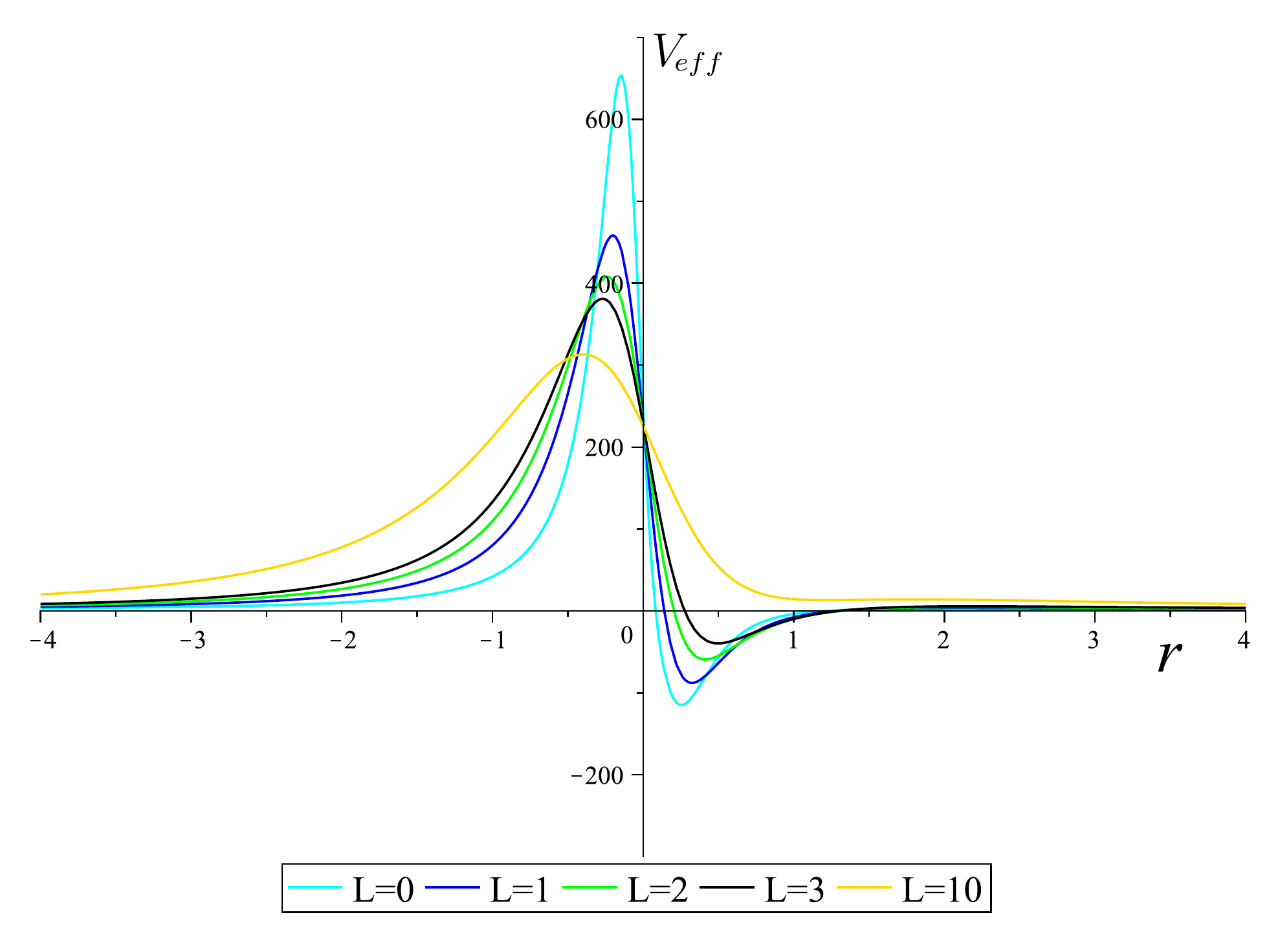}\caption{Plots of
$V_{eff}$ versus $r$ for the spin-0 particles. The physical parameters are
chosen as; $M=m=1,\omega=15,a=0.3,$ and $\lambda=2$.}%
\label{myfig1}%
\end{figure}

From now on, the prime symbol denotes the derivation with respect to $r$. The behavior of the effective potential under the effect of LSB parameter for scalar particles is illustrated in \cref{myfig1}, which shows a significant deduction on the potential peak when the LSB parameter is increased.

\section{Fermionic Perturbations} \label{sec4}

To proceed our analysis with the Dirac fields in the geometry of the Kerr-like black hole, we shall use the four-dimensional Dirac equation formulated in the Newman-Penrose (NP) formalism
\cite{Sara39}. By this way, we aim to derive the effective potentials for the fermionic fields propagating in this geometry. To achieve this goal, we use the orthogonal (dragging) coordinates \cite{Sara37,Sara38} for the metric (\ref{1}) and get

\begin{equation}
ds^{2}=-\frac{\Delta\left(  1+L\right)  }{\Sigma}d\widetilde{\tau}^{2}%
+\frac{\Sigma}{\Delta}dr^{2}+\Sigma d\theta^{2}+\frac{\sin^{2}\theta}{\Sigma
}d\widetilde{\varphi}^{2}, \label{S21}%
\end{equation}
where $\Sigma=\rho^{2}$ and

\begin{equation}
d\widetilde{\tau}^{2}=\left(  dt-a\sqrt{1+L}\sin^{2}\theta d\varphi\right)
^{2}, \label{S22}%
\end{equation}
\begin{equation}
d\widetilde{\varphi}^{2}=\left(  \left(  r^{2}+\left(  1+L\right)
a^{2}\right)  d\varphi-a\sqrt{1+L}dt\right)  ^{2}. \label{S23}%
\end{equation}

The NP tetrad of the Kerr-like black hole geometry can be given by%

\[
l^{\mu}=\frac{1}{\Delta}\left[  \frac{r^{2}+a^{2}\left(  1+L\right)  }%
{\sqrt{1+L}},\Delta,0,a\right]  ,
\]

\[
n^{\mu}=\frac{1}{2\Sigma}\left[  \frac{r^{2}+a^{2}\left(  1+L\right)  }%
{\sqrt{1+L}},-\Delta,0,a\right]  ,
\]

\[
m^{\mu}=\frac{1}{\left(  r+ia\sqrt{1+L}\cos\theta\right)  \sqrt{2}}\left[
ia\sqrt{1+L}\sin\theta,0,1,\frac{i}{\sin\theta}\right]  ,
\]

\begin{equation}
\overline{m}^{\mu}=\frac{1}{\left(  r-ia\sqrt{1+L}\cos\theta\right)  \sqrt{2}%
}\left[  -ia\sqrt{1+L}\sin\theta,0,1,\frac{-i}{\sin\theta}\right]
.\label{is24}%
\end{equation}
where a bar over a quantity denotes complex conjugation. Thus, the dual co-tetrad of Eq. (\ref{is24}) reads

\[
l_{\mu}=\left[  \sqrt{1+L},-\frac{\Sigma}{\Delta},0,-a(1+L)\sin^{2}%
\theta\right]  ,
\]

\[
n_{\mu}=\frac{\Delta}{2\Sigma}\left[  \sqrt{1+L},\frac{\Sigma}{\Delta
},0,-a(1+L)\sin^{2}\theta\right]  ,
\]

\[
m_{\mu}=\frac{1}{\left(  r+ia\sqrt{1+L}\cos\theta\right)  \sqrt{2}}\left[
ia\sqrt{1+L}\sin\theta,0,-\Sigma,-i(r^{2}+a^{2}\left(  1+L\right)  \sin
\theta)\right]  ,
\]

\begin{equation}
\overline{m}_{\mu}=\frac{1}{\left(  r-ia\sqrt{1+L}\cos\theta\right)  \sqrt{2}%
}\left[  -ia\sqrt{1+L}\sin\theta,0,-\Sigma,i(r^{2}+a^{2}\left(  1+L\right)
\sin\theta)\right]  .\label{s25}%
\end{equation}

Before deriving the non-zero spin coefficients, one can re-normalize the NP tetrad by using the spin boost Lorentz transformations:%

\begin{equation}
l\rightarrow \boldsymbol{l}%
%
=\zeta l, \medspace \medspace n\rightarrow\boldsymbol{n}%
%
=\zeta^{-1}n, \label{s26}%
\end{equation}

\begin{equation}
m\rightarrow \boldsymbol{m}%
%
=e^{i\phi}m,\medspace \medspace \overline{m}\rightarrow\boldsymbol{\overline{m}}%
%
=e^{-i\phi}\overline{m},\label{s27}%
\end{equation}
where%

\begin{equation}
\zeta=\sqrt{\frac{\Delta}{2\Sigma}}\text{ \ and \ \ \ }e^{i\phi}=\frac
{\sqrt{\Sigma}}{r-ia\sqrt{1+L}\cos\theta}.\label{s28}%
\end{equation}

Thus, we have
\[
\boldsymbol{l}%
^{\mu}=\frac{1}{\sqrt{2\Delta\Sigma}}\left[  \frac{r^{2}+a^{2}\left(
1+L\right)  }{\sqrt{1+L}},\Delta,0,a\right]  ,
\]

\[
\boldsymbol{n}%
%
^{\mu}=\frac{1}{\sqrt{2\Delta\Sigma}}\left[  \frac{r^{2}+a^{2}\left(
1+L\right)  }{\sqrt{1+L}},-\Delta,0,a\right]  ,
\]

\[
\boldsymbol{m}%
%
^{\mu}=\frac{1}{\sqrt{2\Sigma}}\left[  ia\sqrt{1+L}\sin\theta,0,1,\frac
{i}{\sin\theta}\right]  ,
\]

\begin{equation}
\boldsymbol{\overline{m}}%
\acute{}%
^{\mu}=\frac{1}{\sqrt{2\Sigma}}\left[  -ia\sqrt{1+L}\sin\theta,0,1,\frac
{-i}{\sin\theta}\right]  ,\label{s29}%
\end{equation}
and 

\[
\boldsymbol{l}%
%
_{\mu}=\sqrt{\frac{\Delta}{2\Sigma}}\left[  \sqrt{1+L},-\frac{\Sigma}{\Delta
},0,-a(1+L)\sin^{2}\theta\right],
\]

\[
\boldsymbol{n}%
%
_{\mu}=\sqrt{\frac{\Delta}{2\Sigma}}\left[  \sqrt{1+L},\frac{\Sigma}{\Delta
},0,-a(1+L)\sin^{2}\theta\right],
\]

\[
\boldsymbol{m}%
%
_{\mu}=\frac{1}{\sqrt{2\Sigma}}\left[  ia\sqrt{1+L}\sin\theta,0,-\Sigma
,-i\left(  r^{2}+a^{2}\left(  1+L\right)  \right)  \sin\theta\right],
\]

\begin{equation}
\boldsymbol{\overline{m}}%
%
_{\mu}=\frac{1}{\sqrt{2\Sigma}}\left[  -ia\sqrt{1+L}\sin\theta,0,-\Sigma
,i\left(  r^{2}+a^{2}\left(  1+L\right)  \right)  \sin\theta\right].
\label{s30}%
\end{equation}

The non-zero spin coefficients \cite{29} can then be computed as  

\ \
\[
\pi=-\tau=\frac{\Sigma_{\theta}}{2\Sigma\sqrt{2\Sigma(1+L)}}+i\frac
{a\sin\theta\Sigma_{r}}{2\Sigma\sqrt{2\Sigma}},
\]

\[
\beta=-\alpha=-\frac{\Sigma_{\theta}}{4\Sigma\sqrt{2\Sigma(1+L)}}+\frac
{\cot\theta}{2\sqrt{2\Sigma(1+L)}}-\frac{ia\sin\theta\Sigma_{r}}{4\Sigma
\sqrt{2\Sigma}},
\]

\[
\rho=\mu=-\frac{\Sigma_{r}\sqrt{\Delta}}{2\Sigma\sqrt{2\Sigma}}-i\frac
{a\sqrt{\Delta(1+L)}\cos\theta}{\Sigma\sqrt{2\Sigma}},
\]

\begin{equation}
\varepsilon=\gamma=\frac{\Delta_{r}}{4\sqrt{2\Delta\Sigma}}-\frac{\Delta
\Sigma_{r}}{4\Sigma\sqrt{2\Delta\Sigma}}-i\frac{a\sqrt{\Delta(1+L)}\cos\theta
}{2\Sigma\sqrt{2\Sigma}}. \label{s32}%
\end{equation}

\ \ \ \ \ \ 

After this step, we employ the Chandrasekar-Dirac equations (CDEs) \cite{29} to find the equations governing the fermion fields. CDEs are given by%

\[
\left(  D+\varepsilon-\rho\right)  F_{1}+\left(  \widetilde{\delta}+\pi
-\alpha\right)  F_{2}=i\mu^{\ast}G_{1},
\]

\[
\left(  \Delta+\mu-\gamma\right)  F_{2}+\left(  \delta+\beta-\tau\right)
F_{1}=i\mu^{\ast}G_{2,}%
\]

\[
\left(  D+\overline{\varepsilon}-\overline{\rho}\right)  G_{2}-\left(
\delta+\overline{\pi}-\overline{\alpha}\right)  G_{1}=i\mu^{\ast}F_{2,}%
\]

\begin{equation}
\left(  \Delta+\overline{\mu}-\overline{\gamma}\right)  G_{1}-\left(
\overline{\delta}+\overline{\beta}-\overline{\tau}\right)  G_{2}=i\mu^{\ast
}F_{1},\label{s24}%
\end{equation}
where $F_{1},F_{2},G_{1}$, and $G_{2}$ are the spinor fields and $D,\Delta,\delta,$ and $\overline{\delta}$ are the directional covariant derivative
operators, which are given by%

\begin{equation}
D=\boldsymbol{l}^{\mu}\partial_{\mu},\medspace\Delta=\boldsymbol{n}^{\mu}\partial_{\mu},\medspace\delta=\boldsymbol{m}^{\mu}%
\partial_{\mu},\medspace\overline{\delta}=\boldsymbol{\overline{m}}^{{\mu}}\partial_{\mu}. \label{s25n}%
\end{equation}

The form of the CDEs suggests that 

\[
F_{i}(t,r,\theta,\varphi)=\frac{1}{\sqrt{(r-ia\sqrt{1+L}\cos\theta)}%
}e^{-i\left(  \omega t+m\phi\right)  }\Psi_{i}\left(  r,\theta\right)  ,
\]

\begin{equation}
G_{i}\left(  t,r,\theta,\phi\right)  =\frac{1}{\sqrt{(r+ia\sqrt{1+L}\cos
\theta)}}e^{-i\left(  \omega t+m\phi\right)  }\Phi_{i}\left(  r,\theta\right)
. \label{s33}%
\end{equation}

Inserting \cref{s29,s30,s32,s25n,s33} into the CDEs (\ref{s24}), we obtain

\begin{multline}
\left\{  \sqrt{\Delta}\partial_{r}-\frac{i\omega\left(  r^{2}+\left(
1+L\right)  a^{2}\right)  }{\sqrt{\Delta\left(  1+L\right)  }}+\frac
{\Delta_{r}}{4\sqrt{\Delta}}-\frac{ima}{\sqrt{\Delta}}\right\}  \Psi
_{1}\left(  r,\theta\right)  +\\
\frac{1}{\sqrt{1+L}}\left\{  \partial_{\theta}-\frac{m}{\sin\theta}%
-a\omega\sqrt{1+L}\sin\theta+\frac{\cot\theta}{2}\right\}  \Psi_{2}\left(
r,\theta\right)  =\\
i\mu\left(  r-ia\sqrt{1+L}\cos\theta\right)  \Phi_{1}\left(  r,\theta\right)
, \label{K1}%
\end{multline}

\begin{multline}
-\left\{  \sqrt{\Delta}\partial_{r}+\frac{i\omega\left(  r^{2}+\left(
1+L\right)  a^{2}\right)  }{\sqrt{\Delta\left(  1+L\right)  }}+\frac
{\Delta_{r}}{4\sqrt{\Delta}}+\frac{ima}{\sqrt{\Delta}}\right\}  \Psi
_{2}\left(  r,\theta\right)  +\\
\frac{1}{\sqrt{1+L}}\left\{  \partial_{\theta}-\frac{m}{\sin\theta}%
-a\omega\sqrt{1+L}\sin\theta+\frac{\cot\theta}{2}\right\}  \Psi_{1}\left(
r,\theta\right)  =\\
i\mu\left(  r-ia\sqrt{1+L}\cos\theta\right)  \Phi_{2}\left(  r,\theta\right)
, \label{K2}%
\end{multline}

\begin{multline}
\left\{  \sqrt{\Delta}\partial_{r}-\frac{i\omega\left(  r^{2}+\left(
1+L\right)  a^{2}\right)  }{\sqrt{\Delta\left(  1+L\right)  }}+\frac
{\Delta_{r}}{4\sqrt{\Delta}}-\frac{ima}{\sqrt{\Delta}}\right\}  \Phi
_{2}\left(  r,\theta\right)  -\\
\frac{1}{\sqrt{1+L}}\left\{  \partial_{\theta}+\frac{m}{\sin\theta}%
+a\omega\sqrt{1+L}\sin\theta+\frac{\cot\theta}{2}\right\}  \Phi_{1}\left(
r,\theta\right)  =\\
i\mu\left(  r+ia\sqrt{1+L}\cos\theta\right)  \Psi_{2}\left(  r,\theta\right)
, \label{K3}%
\end{multline}

\begin{multline}
-\left\{  \sqrt{\Delta}\partial_{r}+\frac{i\omega\left(  r^{2}+\left(
1+L\right)  a^{2}\right)  }{\sqrt{\Delta\left(  1+L\right)  }}+\frac
{\Delta_{r}}{4\sqrt{\Delta}}+\frac{ima}{\sqrt{\Delta}}\right\}  \Phi
_{1}\left(  r,\theta\right)  -\\
\frac{1}{\sqrt{1+L}}\left\{  \partial_{\theta}-\frac{m}{\sin\theta}%
-a\omega\sqrt{1+L}\sin\theta+\frac{\cot\theta}{2}\right\}  \Phi_{2}\left(
r,\theta\right)  =\\
i\mu\left(  r+ia\sqrt{1+L}\cos\theta\right)  \Psi_{1}\left(  r,\theta\right)
. \label{K4}%
\end{multline}

Since the functions $\Psi_{i}\left(  r,\theta\right)$ and $\Phi_{i}\left(  r,\theta\right)$ depend on the radial and angular variables, one can separate them by the following ansatzes%
\begin{align}
\Psi_{1}\left(  r,\theta\right)   &  =\Re_{+}\left(  r\right)  \aleph
_{+}\left(  \theta\right)  ,\\
\Psi_{2}\left(  r,\theta\right)   &  =\Re_{-}\left(  r\right)  \aleph
_{-}\left(  \theta\right)  ,\label{S36}%
\end{align}
where $\Re_{\pm}\left(  r\right)  =\Delta^{-1/4}p(r)_{\pm1/2}$. Using the tortoise coordinate $\left(  r_{\ast}\right)  $ as
$\frac{d}{dr_{\ast}}=\frac{\Delta\sqrt{1+L}}{r^{2}+a^{2}\left(  1+L\right)
}\frac{d}{dr}$, \cref{K1,K2,K3,K4} yield the following two one-dimensional
Schr\"{o}dinger-like radial equations:%

\begin{align}
\left\{  \frac{d}{dr_{\ast}}-i\varpi\right\}  p_{+1/2} &  =\frac{\lambda
}{\sqrt{1+L}}\frac{\sqrt{\Delta}}{K}p_{-1/2},\label{Sk37}\\
-\left\{  \frac{d}{dr_{\ast}}+i\varpi\right\}  p_{-1/2} &  =\frac{\lambda
}{\sqrt{1+L}}\frac{\sqrt{\Delta}}{K}p_{+1/2},\label{S37}%
\end{align}
in which%

\begin{equation}
K=\frac{r^{2}+a^{2}(1+L)}{\sqrt{1+L}}, \quad \varpi=\omega+\frac{ma}%
{K}. \label{SS37}%
\end{equation}

Setting the eigenvalue $\lambda=-\left(  l+\frac{1}{2}\right)$   for the angular equations as%

\begin{equation}
\frac{%
\mathcal{L}%
^{\dagger}\aleph_{-}\left(  \theta\right)  }{\aleph_{+}\left(  \theta\right)
}=-\lambda,\text{ \ \ \ \ \ \ \ }\frac{%
\mathcal{L}%
\aleph_{+}\left(  \theta\right)  }{\aleph_{-}\left(  \theta\right)  }%
=\lambda, \label{SK37}%
\end{equation}
where $\mathcal{L}$ and $\mathcal{L}^{\dagger}$ are the angular operators%

\begin{align}
\mathcal{L}
&  =\partial_{\theta}+\frac{m}{\sin\theta}+\frac{\cot\theta}{2}+a\omega
\sqrt{1+L}\sin\theta,\text{ \ }\\
\text{\ \ \ \ }%
\mathcal{L}%
^{\dagger} &  =\partial_{\theta}-\frac{m}{\sin\theta}+\frac{\cot\theta}%
{2}-a\omega\sqrt{1+L}\sin\theta,\label{SS38}%
\end{align}
one can have the spin-weighted spherical harmonics \cite{spinw} for the angular equations. Moreover, if we let

\begin{equation}
Z_{+}=p_{+1/2}+p_{-1/2},\text{ \ \ \ \ }Z_{-}=p_{-1/2}-p_{+1/2}, \label{S41}%
\end{equation}
equations (\ref{Sk37}) and (\ref{S37}) can be transformed to one-dimensional Schr\"{o}dinger-like
wave equations:%

\begin{equation}
\left(  \frac{d^{2}}{dr_{\ast}^{2}}+\varpi^{2}\right)  Z_{\pm}=V_{eff}^{\pm}Z_{\pm
}.\label{S42}%
\end{equation}

From now on, for the sake of simplicity, we consider the massless ($\mu=0$) fermions. In this case, the effective potentials for the propagating Dirac fields become%

\begin{equation}
V_{eff}^{\pm}=\frac{\Delta}{K}\left\{  \frac{\lambda^{2}}{K\left(  1+L\right)  }%
\pm\frac{d}{dr}\left(  \frac{\lambda\sqrt{\Delta}}{K}\right)  \right\}
.\label{S43}%
\end{equation}

\begin{figure}[h]
\centering\includegraphics[scale=0.6]{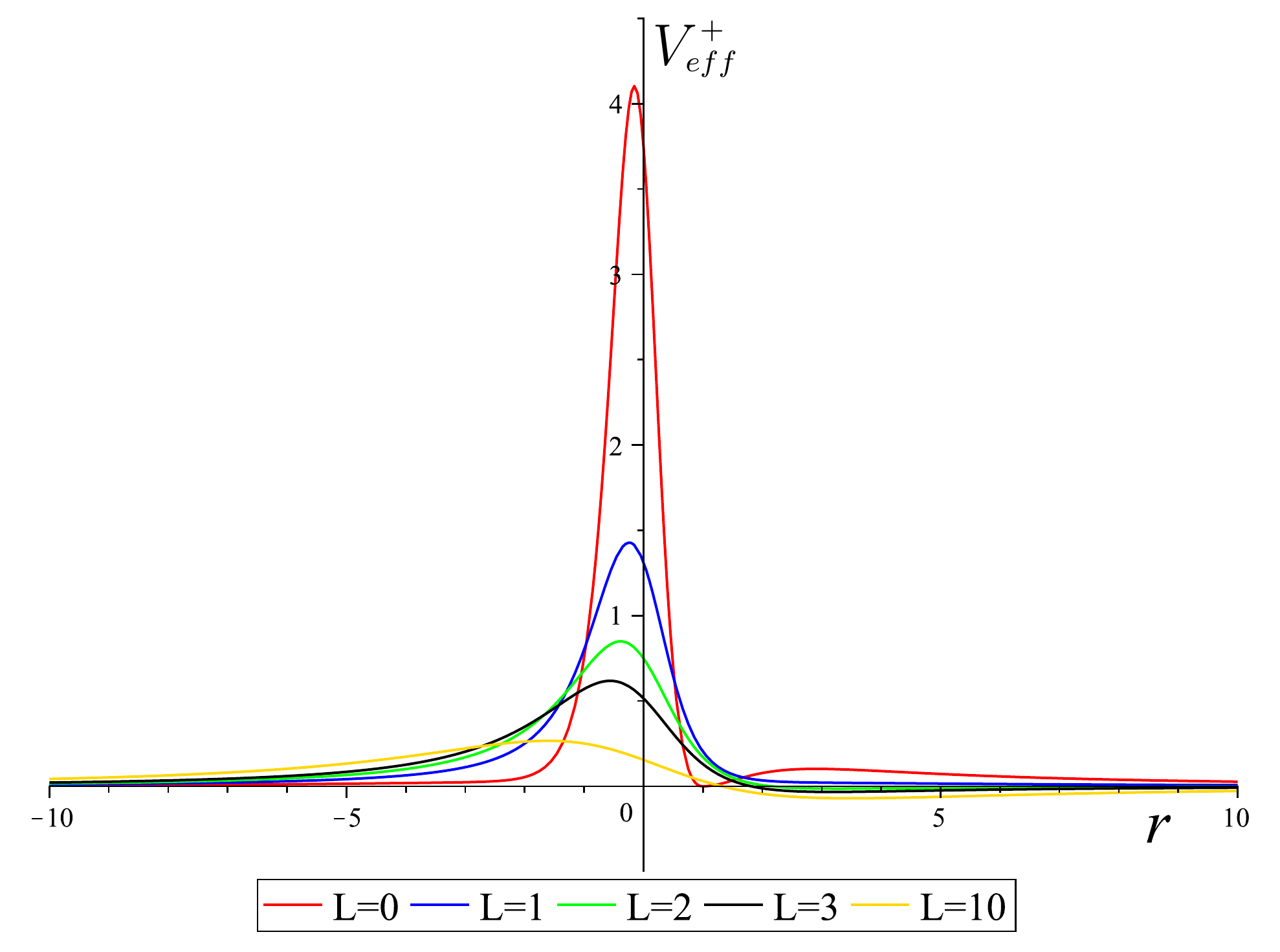}\caption{Plots of
$V_{eff}^{+}$ versus $r$ for the spin-($\nicefrac{+1}{2}$) particles.  The physical
parameters are chosen as; $M=a=1,$ and $\lambda=-1.5$. } \label{myfig2}
\end{figure}

\begin{figure}[h]
\centering\includegraphics[scale=0.6]{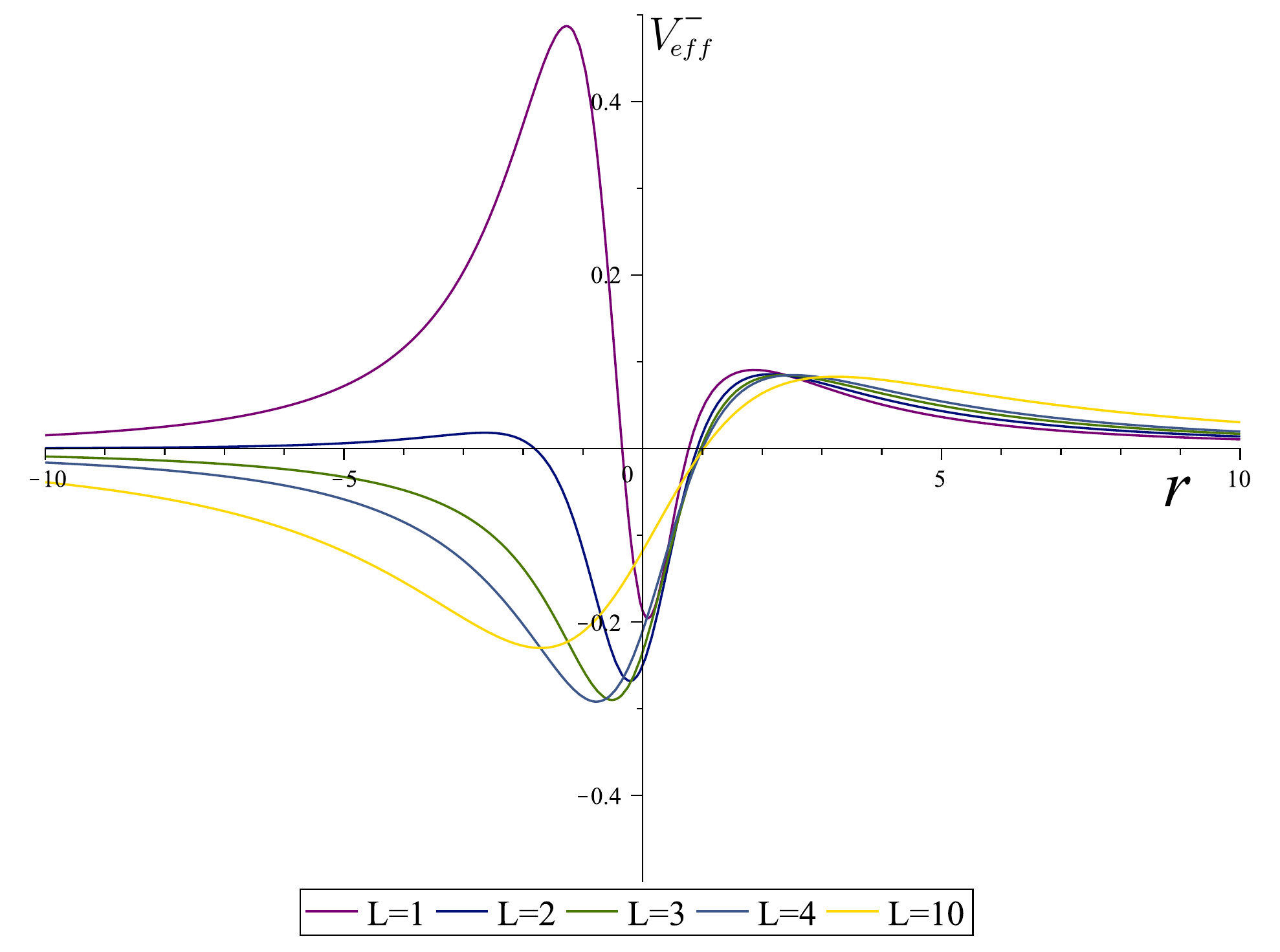}\caption{Plots of
$V_{eff}^{-}$ versus $r$ for fermions spin-($\nicefrac{-1}{2}$) particles. The physical
parameters are chosen as; $M=a=1,$ and $\lambda=-1.5$.}%
\label{myfig3}
\end{figure}

The behaviors of the effective potentials (\ref{S43}) are depicted in \cref{myfig2,myfig3}, which stand for spin-up and spin-down particles, respectively.

\section{Greybody radiation in Kerr spacetime of BGM} \label{sec5}

\subsection{GFs of Bosons}

Studying GFs provides important clues on the quantum structure of black holes. Derivation of the GF can be conducted as
(\cite{Sara33})%
\begin{equation}
\sigma_{\ell}\left(  \omega\right)  \geq\sec h^{2}\left(  \int_{-\infty
}^{+\infty}\wp dr_{\ast}\right)  ,\label{21}%
\end{equation}
where
\begin{equation}
\wp=\frac{\sqrt{\left(  h%
\acute{}%
\right)  ^{2}+\left(  \omega^{2}-V_{eff}-h^{2}\right)  ^{2}}}{2h}.\label{22}%
\end{equation}

In addition, the following conditions must be met: 1) $h\left(  r_{\ast}\right)
\succ0$ and 2) $h\left(  -\infty\right)  =h\left(  \infty\right)  =\omega$. Without loss of generality, if one simply sets $h=\omega,$ thus GF formula (\ref{21}) reduces to
\begin{equation}
\sigma_{\ell}\left(  \omega\right)  \geq\sec h^{2}\left(  \int_{-\infty
}^{+\infty}\frac{V_{eff}}{2\omega}dr_{\ast}\right)  \text{.}\label{23}%
\end{equation}

To have an immaculately integration, let us consider the massless form of the bosonic
effective potential (\ref{20}). Since the tortoise coordinate (\ref{17}) varies from $-\infty$ (the event horizon $r_{h}$: lower boundary of the integral) and to $+\infty$ (spatial infinity: upper boundary of the integral) in Eq. (\ref{23}), we get%
\begin{multline}
\sigma_{\ell}\left(  \omega\right)  \geq\sec h^{2}\left(  \frac{1}{2\omega
}\int_{r_{h}}^{+\infty}\frac{\sqrt{1+L}}{\left(  r^{2}+\left(  1+L\right)
a^{2}\right)  }\left[  \frac{\Delta%
\acute{}%
r}{\left(  r^{2}+\left(  1+L\right)  a^{2}\right)  }+\right.  \right.
\label{24}\\
\left.  \frac{\Delta}{\left(  r^{2}+\left(  1+L\right)  a^{2}\right)  }%
-\frac{3r^{2}\Delta}{\left(  r^{2}+\left(  1+L\right)  a^{2}\right)  ^{2}%
}+\frac{4Mram\omega}{\Delta\sqrt{1+L}}\right.  \\
\left.  \left.  -\frac{m^{2}a^{2}}{\Delta}+\left(  \omega^{2}a^{2}\left(
1+L\right)  +\lambda\right)  \right]  dr\right). 
\end{multline}\label{D24}

After using the series and evaluating the integral, the GF can be obtained as follows%
\begin{multline}
\sigma_{\ell}\left(  \omega\right)  \geq\sec h^{2}\left(  \frac{\sqrt{1+L}%
}{2\omega}\right)  \left\{  \left(  \frac{-8a^{2}}{3r_{h}^{3}}+\frac{M}%
{r_{h}^{2}\left(  1+L\right)  }-\frac{2a^{4}\left(  1+L\right)  }{5r_{h}^{5}%
}+\frac{2Ma^{2}}{r_{h}^{4}}\right)  -\right.  \label{25}\\
\left.  m^{2}a^{2}\left(  1+L\right)  \left(  \frac{1}{3r_{h}^{3}}+\frac
{M}{2r_{h}^{4}}+\frac{4M^{2}-2a^{2}\left(  1+L\right)  }{5r_{h}^{5}}%
-\frac{3Ma^{2}\left(  1+L\right)  -4M^{3}}{3r_{h}^{6}}\right)  +\right.  \\
\left.  \left(  \omega^{2}a^{2}\left(  1+L\right)  +\lambda\right)  \left[
\frac{1}{r_{h}}-\frac{\left(  1+L\right)  a^{2}}{3r_{h}^{3}}+\frac{\left(
1+L\right)  ^{2}a^{4}}{5r_{h}^{5}}\right]  \right.  \\
\left.  +4Mam\omega\sqrt{1+L}\left(  \frac{1}{2r_{h}^{2}}+\frac{2M}{3r_{h}%
^{3}}-\frac{\left(  1+L\right)  a^{2}-2M^{2}}{2r_{h}^{4}}-\frac{6Ma^{2}\left(
1+L\right)  -8M^{3}}{5r_{h}^{5}}\right)  \right\}. 
\end{multline}\label{D25}

\begin{figure}[h]
\centering\includegraphics[scale=0.6]{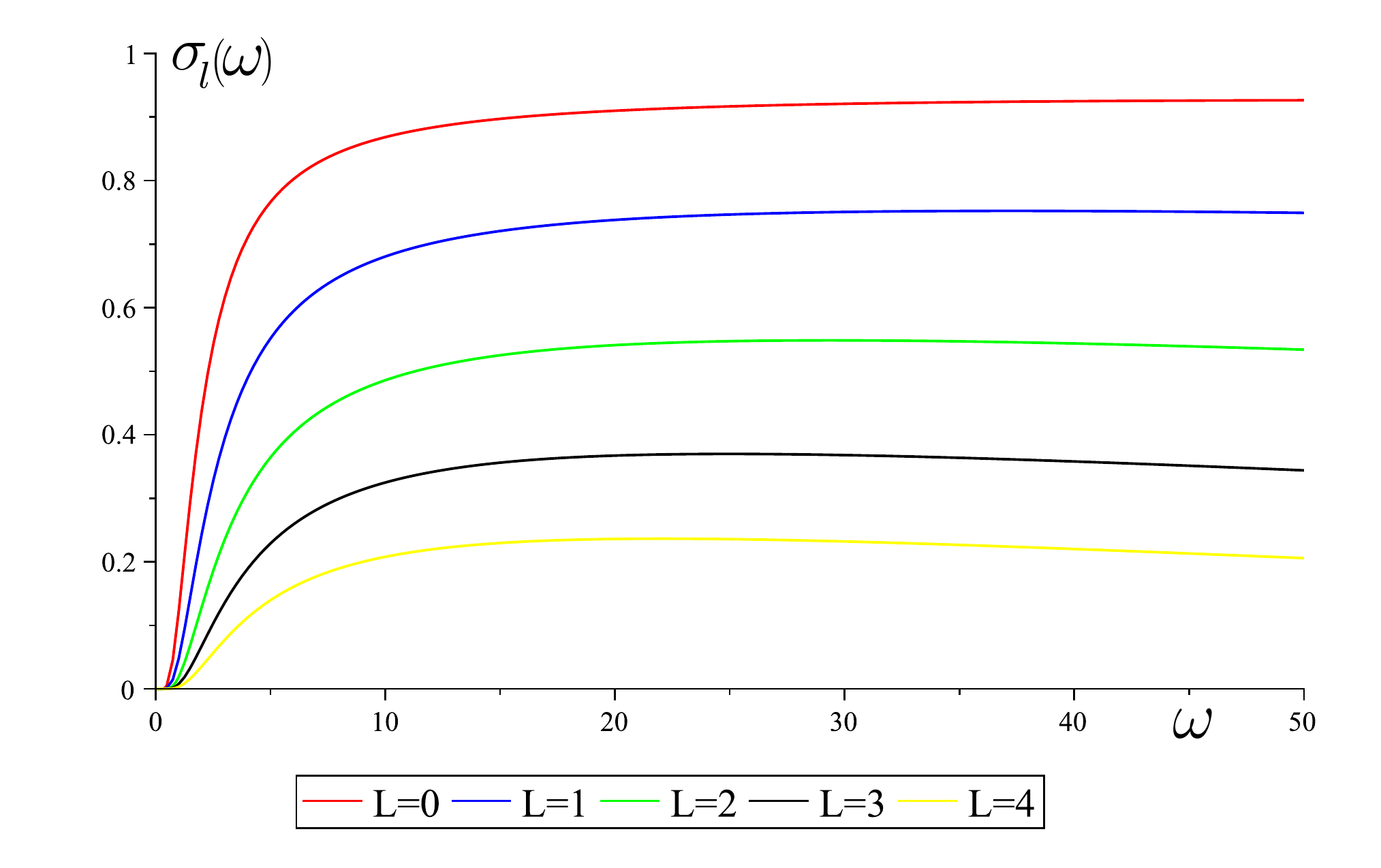}\caption{Plots of
$\sigma_{l}\left(  \omega\right)  $ versus $\omega$ for the spin-0 particles.
The physical parameters are chosen as; $M=r=1,a=0.03$ and $\lambda=2$.} \label{myfig4}
\end{figure}

In \cref{myfig4}, the behaviors of the obtained bosonic GF of the
Kerr-like black hole are demonstrated. Thus, the effect of LSB on the bosonic GF is visualized. As can be seen from the plots, $\sigma_{\ell}$ clearly decreases with the increasing LSB parameter. Namely, LSB plays a kind of
fortifier role for the GF of spin-$0$ particles.

\subsection{GFs of Fermions}

In this section, we shall concentrate on the GF of the fermions to
elicit the effect of the LSB on their emission from the Kerr-like black hole in the BGM. For this purpose, we use the effective potentials (\ref{S43}) in Eq. (\ref{23}):

\begin{equation}
\sigma_{l}\left(  \omega\right)  \succeq\sec h^{2}\left(  \frac{1}{2\omega
}\int_{r_{h}}^{+\infty}\left\{  \frac{\lambda^{2}dr}{\left(  r^{2}%
+a^{2}\left(  1+L\right)  \right)  \sqrt{1+L}}\pm\lambda\frac{d}{dr}\left(
\frac{\sqrt{\Delta\left(  1+L\right)  }}{r^{2}+a^{2}\left(  1+L\right)
}\right)  \right\}  dr\right). \label{SK26}%
\end{equation}

We then employ the classical term-by-term integration technique used for obtaining asymptotic expansions of integral, which requires the integrand to have an uniform asymptotic expansion in the integration variable \cite{asym}. Thus, the evaluation of the integral (\ref{SK26}) yields

\begin{multline}
\sigma_{l\left(  \pm\right)  }\left(  \omega\right)  \succeq\sec h^{2}\left(
\frac{\lambda}{2\omega}\left\{  \frac{\lambda}{\sqrt{1+L}r_{h}}\left(
1-\frac{a^{2}\left(  1+L\right)  }{3r_{h}^{2}}\right)  \pm\left(  \frac
{M^{2}-a^{2}\left(  1+L\right)  }{2r_{h}^{3}}\right)  \right.  \right.  \\
\left.  \left.  \mp\left[  M\left(  \frac{M}{3r_{h}^{3}}+\frac{3}{8}%
\frac{M^{2}-a^{2}\left(  1+L\right)  }{r_{h}^{4}}\right)  +\frac{1}{r_{h}%
}-\frac{M}{r_{h}^{2}}\right]  \right\}  \right)  .\label{SK27}%
\end{multline}

\begin{figure}[h]
\centering\includegraphics[scale=0.6]{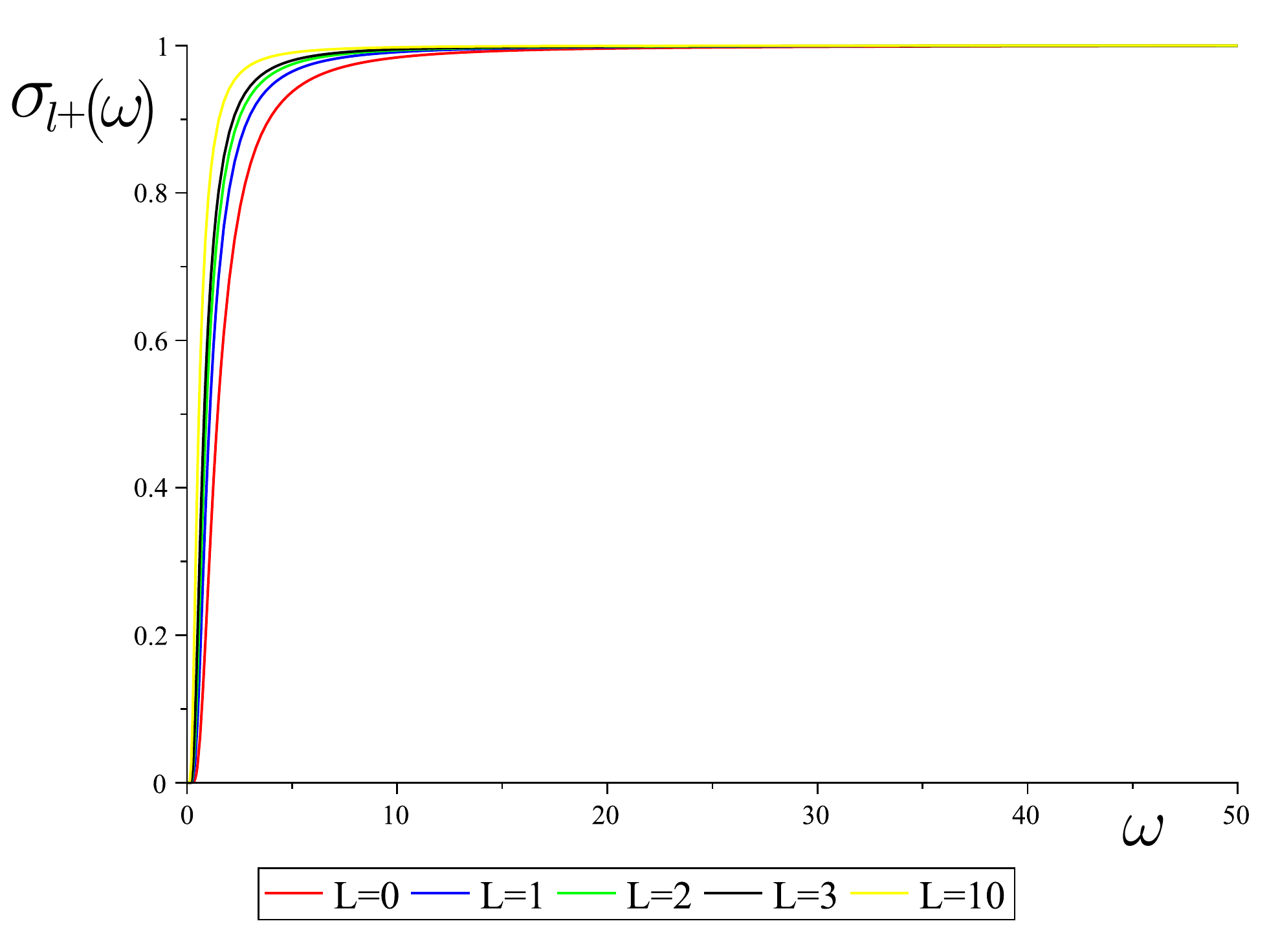}\caption{Plots of
$\sigma_{l+}\left(  \omega\right)  $ versus $\omega$ for fermions with spin-($\nicefrac{+1}{2}$). The physical parameters are chosen as; $M=r=1,a=0.03$ and
$\lambda=-1.5$.} \label{myfig5}
\end{figure}

\begin{figure}[h]
\centering\includegraphics[scale=0.6]{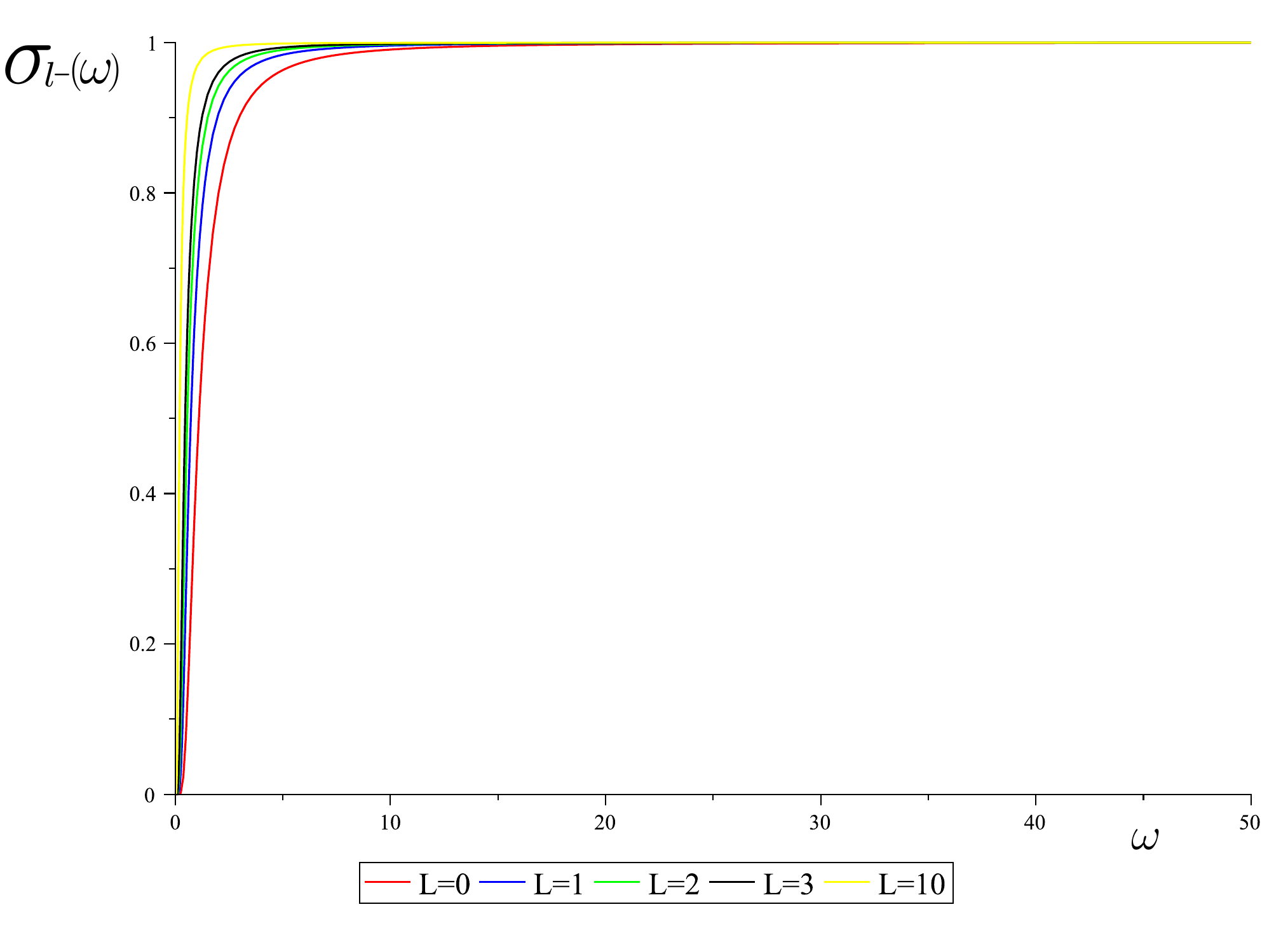}\caption{Plots of
$\sigma_{l-}\left(  \omega\right)  $ versus $\omega$ for the spin-($\nicefrac{-1}{2}$) particles. The physical parameters are chosen as; $M=r=1,a=0.03$ and
$\lambda=-1.5$.} \label{myfig6}
\end{figure}

The behaviors of both spin-($\nicefrac{+1}{2}$) and spin-($\nicefrac{-1}{2}$) under the influence of LSB effect are
depicted in \cref{myfig5,myfig6}, respectively. 

\section{Effect of LSB Parameter on QNM\lowercase{s}} \label{sec6}
Non-trivial information about thermalisation in quantum field theory are obtained by studying small perturbations of a black hole away form the equilibrium. Such perturbations are described by QNMs  \cite{Sarak34,Sarak35}. These special oscillations are similar to normal modes of a closed system. However, since the perturbation can
fall into the black hole or radiate to infinity, the corresponding frequencies are complex \cite{Sarak36}. The oscillation frequency is defined by the real part and the rate of specific damping mode as a result of a radiation is determined by imaginary
part. Thus, by getting the QNMs in the BGM, the comparison of theoretical predictions with the experimental data supplied by "future" LIGO and VIRGO type experiments would help us to numerically constrain the LSB parameter. Thus, in general, it is important to accumulate data on QNMs of black holes in various theories of gravity \cite{data}.

In this section, for both scalar and Dirac perturbations, we follow the WKB approximation method \cite{Sara34,Sara35} to derive the frequencies of the QNMs of the Kerr-like black hole in the BGM. To this end, we shall simply transform the obtained one-dimensional Schr\"odinger-like equations [ see Eqs. (\ref{19}) and (\ref{S42})] to the following Zerilli \cite{29} type differential equations:

\begin{equation}
\frac{d^2 Z}{dr^{*2}}+ V_{Geff}Z=0,  \label{isch}  
\end{equation}
where $Z$ function is assumed to have a time-dependence $e^{i \omega t}$, $V_{Geff}$ is the generic effective potential, and $r_{*}$ is the tortoise coordinate as being stated above.

\subsection{Scalar QNMs}
Comparing with the numerical results, the WKB approach  \cite {wkb1} is known to lead to good predictions for obtaining the QNMs. In this method, $V_{Geff}$  is written by using the tortoise coordinate so as to be constant at $r_{\ast}\rightarrow 0$ (event horizon) and at $r_{\ast}\rightarrow + \infty $ (spatial infinity). The maximum value of $V_{Geff}$, which we symbolize it as $V_{0}$ from now on, is achieved at $r^0_{*}$. Three regions are defined as follows: region-$I$ from $ - \infty$ to $r_1$, which is the first turning point where the potential becomes zero, region-$II$ from $r_1$ to $r_2$, namely the second turning point, and region-$III$ from $r_2$ to $+\infty$. In region-$II$, the Taylor expansion is made over $r^0_{\ast}$. In regions-$I$ and -$III$, the solution can be approximated by an exponential function:

\begin{equation}
    Z \sim exp \bigg[ \frac{1}{\zeta}\sum _{n=0} ^{\infty} \zeta ^n \Xi_n(x) \bigg],  \quad \quad \zeta \rightarrow 0.
\end{equation}

This expression can be substituted in Eq. (\ref{isch}) to get $\Xi_n$ as a function of the potential and its derivative. We then impose the boundary conditions of the QNMs:
\begin{eqnarray}
Z \sim e^{-i \omega r_{\ast}} \ \ \ \ \ \ r_{\ast} \rightarrow - \infty, \\
Z \sim e^{i \omega r_{\ast}} \ \ \ \ \ \ r_{\ast} \rightarrow + \infty, \label{isBC}
\end{eqnarray} 
and match the solutions of regions-$I$ and -$III$ with the solution of region-$II$ at the turning points, $r_1$ and $r_2$, respectively. The WKB approximation can be extended from the third to sixth order. This allows us to obtain the complex frequencies of the QNMs from the following expression \cite{wkb1}:

\begin{equation}
\omega^{2}=\left[  V_{0}+\sqrt{2V_{0}^{%
\acute{}%
\acute{}%
}}\Lambda\left(  n\right)  -i\left(  n+\frac{1}{2}\right)  \sqrt{-2V_{0}^{%
\acute{}%
\acute{}%
}}\left(  1+\Omega\left(  n\right)  \right)  \right]  ,\label{26}%
\end{equation}
where%

\begin{equation}
\Lambda\left(  n\right)  =\frac{1}{\sqrt{-2V_{0}^{%
\acute{}%
\acute{}%
}}}\left[  \frac{1}{8}\left(  \frac{V_{0}^{\left(  4\right)  }}{V_{0}^{%
\acute{}%
\acute{}%
}}\right)  \left(  \frac{1}{4}+\alpha^{2}\right)  -\frac{1}{288}\left(
\frac{V_{0}^{%
\acute{}%
\acute{}%
\acute{}%
}}{V_{0}^{%
\acute{}%
\acute{}%
}}\right)  ^{2}\left(  7+60\alpha^{2}\right)  \right]  ,\label{27}%
\end{equation}
and%

\begin{multline}
\Omega\left(  n\right)  =\frac{1}{-2V_{0}^{%
\acute{}%
\acute{}%
}}\left[  \frac{5}{6912}\left(  \frac{V_{0}^{%
\acute{}%
\acute{}%
\acute{}%
}}{V_{0}^{%
\acute{}%
\acute{}%
}}\right)  ^{4}\left(  77+188\alpha^{2}\right)  -\frac{1}{384}\left(
\frac{V_{0}^{%
\acute{}%
\acute{}%
\acute{}%
2}V_{0}^{\left(  4\right)  }}{V_{0}^{%
\acute{}%
\acute{}%
3}}\right)  \left(  51+100\alpha^{2}\right)  +\right.  \label{28}\\
\left.  \frac{1}{2304}\left(  \frac{V_{0}^{\left(  4\right)  }}{V_{0}^{%
\acute{}%
\acute{}%
}}\right)  ^{2}\left(  67+68\alpha^{2}\right)  +\frac{1}{288}\left(
\frac{V_{0}^{%
\acute{}%
\acute{}%
\acute{}%
}V_{0}^{\left(  5\right)  }}{V_{0}^{%
\acute{}%
\acute{}%
2}}\right)  \left(  19+28\alpha^{2}\right)  -\frac{1}{288}\left(  \frac
{V_{0}^{\left(  6\right)  }}{V_{0}^{%
\acute{}%
\acute{}%
}}\right)  \left(  5+4\alpha^{2}\right)  \right]  ,
\end{multline}
where the prime symbol denotes the differentiation with respect to $r_{\ast}$. The value of $r^0_{\ast}$ is determined, and $\alpha=n+\frac{1}{2}$. With the help of the effective potential (\ref{20}), one can easily get $V_{Geff}$. After making straightforward calculations and numerical analysis, we have obtained the bosonic QNMs, which are tabulated in \cref{tab1} for the angular momentum $l=2$. In \cref{tab1}, the case of $m=0$ for the first (fundamental) overtone $n=0$ is considered (higher tones also show similar results that are parallel to the behaviors of the $n=0$ mode). The revealed knowledge from \cref{tab1} is that the oscillations decrease when the LSB parameter increases.  But for the damping rate, there is no ostensive information about the LSB effect. At the beginning ($L	\approx0-1.2$), we have a decrease in the imaginary part of the QNM frequencies but then this linear relationship between them disappears and shows an irregular behavior. Those bizarre behaviors can be understandable from \cref{myfig1} which shows the potential barriers that scalar QNMs are affected: the potential can take negative and positive values.
 
 \subsection{Dirac QNMs}
In this sub-section, we shall apply the methodology applied for QNM bosons in the previous section to the fermions. To this end, we consider the potentials obtained in Eq. (\ref{S43}) and use them in \cref{26,27,28}. \cref{tab2} constitutes the main results of this part: the numerically computed QNM frequencies for varying values of the LSB parameter for the fix rotating parameter $a=0.4$ are displayed in the table. It is obvious from \cref{tab2} that both oscillatory and damping parts of the fermionic QNMs tend to decrease with the increasing LSB parameter. On the contrary, they increase with increasing $l$ and $n$ values: hence, the characteristic fermionic QNMs are different from the bosonic ones.

\begin{minipage}[c]{.40\textwidth}
   \centering
  \begin{tabular}{ |c|c| }
\hline
$L$ & $\omega_{bosons}$ \\
\hline
 0 &  0.374411-0.089694i\\
 1 &  0.372640-0.045766i \\
1.1 & 0.372553-0.044986i \\
1.2 & 0.372473-0.044784i \\
1.3 & 0.372399-0.045302i \\
1.4 & 0.372332-0.046902i \\
1.5 & 0.372269-0.050064i \\
1.6 & 0.372210-0.056650i \\
1.7 & 0.372155-0.073158i \\
1.8 & 0.372104-0.244808i \\
1.9 & 0.372063-0.099344i\\
2   & 0.372025-0.071826i\\
2.1 & 0.371990-0.058375i \\
2.2 & 0.371956-0.050093i\\ 
\hline 
\end{tabular}
\captionof{table}{QNMs of scalar waves in the Kerr-like black hole spacetime.}\label{tab1}
\end{minipage}\qquad
\begin{minipage}[c]{.40\textwidth}
   \centering
   \begin{tabular}{ |c|c|c|c|}
\hline
$l$ & $n$ & $L$ & $\omega_{fermions}$ \\
\hline\hline
1 & 0 & 1 & 0.228468-0.065297i\\
  &   & 2 & 0.191896-0.052521i \\
  &   & 3 & 0.170421-0.044723i \\
\hline
1 & 1 & 1 & 0.265474-0.169833i \\
  &   & 2 & 0.221535-0.137485i \\
  &   & 3 & 0.195343-0.117735i \\
\hline
2 & 0 & 1 & 0.360666-0.065766i \\
  &   & 2 & 0.300382-0.052794i \\
  &   & 3 & 0.265097-0.044763i \\
\hline
  & 1 & 1 & 0.389890-0.182970i\\
  &   & 2 & 0.323537-0.147340i\\
  &   & 3 & 0.284319-0.125374i \\
\hline
  & 2 & 1 & 0.427069-0.279806i\\
  &   & 2 & 0.353472-0.225666i\\
  &   & 3 & 0.309573-0.192416i \\
\hline
\end{tabular}
   \captionof{table}{QNMs of Dirac waves in the Kerr-like black hole spacetime.} \label{tab2}
\end{minipage}

\section{Conclusion} \label{sec7}

Very recently, it has been shown \cite{18} that Kerr-like black hole solutions are existed in the BGM. For the first time with this study, we have studied the GFs and QNMs of the scalar an fermionic fields for the asymptotically flat black holes in Kerr-like black hole. To this end, we have studied the perturbations of the scalar and Dirac fields, respectively. We have summarized the results of our study in \cref{tab1,tab2}.

We have computed GFs for both spin-{0} and spin-($\pm\nicefrac{1}{2}$) particles. As a result of our analysis, we have seen that while the increase in the LSB parameter for scalar waves decreases the GFs, however for the fermionic waves, by making the opposite effect, the increases in the LSB cause also the increment in the GFs. Those remarkable behaviors are clearly depicted in \cref{myfig4,myfig5,myfig6}. Moreover, for the bosonic QNMs, the oscillations decrease when the LSB parameter increases.  But for the damping rate, there is no palpable behavior about the LSB effect. However, these results make sense when looking at the plots of the potential \cref{myfig1}, takes negative and positive values. For the fermionic case, we have deduced from \cref{tab2} that the fermionic QNMs (both the oscillatory and damping parts of the complex frequency) tend to decrease once the LSB paramete increases. On the contrary, QNM frequencies increase with the increment of $l$ and $n$ values: hence, the spin-($\pm\nicefrac{1}{2}$) QNMs exhibit different character compared with the spin-{0} ones. Beyon all those, one may require to perform a full-time domain analysis in order to understand the complete stability feature of the spacetime under the perturbations. However, the present study therefore can only give the qualitative nature of variations of the QNM frequencies with the LSB parameter as far as the bosonic/fermionic perturbations are concerned.

One of the most powerful uses of BGMs is to potentially explain dark energy, which is the phenomenon responsible for the observed accelerated expansion of the universe. Therefore, the GF/QNM analyses of the black hole in the AdS background within the framework of the BGM will be an important future extension of the present work.  This may also be important to understand the AdS/CFT conjecture with LIV, since QNMs are responsible from the stability in the CFT side. This is the next stage of study that interests us.


\begin{thebibliography}{99}                                                                                               %


\bibitem {1}D. Mattingly, Living Rev. Relat. \textbf{8}, 5 (2005).

\bibitem{IS1}
S.~Liberati and L.~Maccione,
Ann. Rev. Nucl. Part. Sci. \textbf{59}, 245-267 (2009).

\bibitem {2}O. Bertolami, J. P\'{a}ramos, Phys. Rev. D \textbf{72}, 044001 (2005).

\bibitem {3}V. A. Kostelecky, S. Samuel, Phys. Rev. D \textbf{39}, 683 (1989);
Phys. Rev. Lett.\textbf{66}, 1811 (1991).

\bibitem {4}V. A. Kostelecky, R. Potting, Phys. Rev. D \textbf{51}, 3923 (1995).

\bibitem{IS2} H. B. Nielsen and M. Ninomiya, Nucl. Phys. B \textbf{141}, 153 (1978).

\bibitem{IS3}J. Ellis, M. K. Gaillard, D. V. Nanopoulos, and S. Rudaz, Nucl. Phys. B \textbf{176}, 61 (1980).

\bibitem{IS4} H. B. Nielsen and I. Picek, Nucl. Phys. B \textbf{211}, 269 (1983).


\bibitem {5}J. Nishimura, G. Vernizzi, JHEP \textbf{04} (2000).

\bibitem {6}R. Gambini, J. Pullin, Phys.Rev. D \textbf{59, }124021\textbf{\ }(1999).

\bibitem {7}S.R. Coleman, S.L. Glashow, Phys. Rev. D \textbf{59}, 116008 (1999).

\bibitem {8}D. Colladay, V. A. Kosteleck\'{y}, Phys. Rev. D \textbf{58},
116002 (1998).

\bibitem {9}D. Colladay, V. A. Kosteleck\'{y}, Phys. Rev. D \textbf{55}, 6760
(1997). \ \ \ \ \ \ 

\bibitem {10}V. A. Kosteleck\'{y}, Phys. Rev. D \textbf{69}, 105009 (2004).\ \ \ \ 

\bibitem {11}R. Bluhm, N.L. Gagne, R. Potting, A. Vrublevskis, Phys. Rev. D
\textbf{77}, 125007 (2008)

\bibitem {12}J.W. Moffat, Intl. J. Mod. Phys. D \textbf{12}, 1279 (2003).

\bibitem {13}S.M. Carroll and E.A. Lim, Phys. Rev. D \textbf{70}, 123525(2004).

\bibitem {14}V.A. Kosteleck\'{y}, S. Samuel, Phys. Rev. D \textbf{40}, 1886 (1989).

\bibitem {s14}M.H. Dickinson, F.O. Lehmann, S.P. Sane, Science \textbf{284},
1954 (1999).

\bibitem {s15}J. P%
\'{}%
aramos, G. Guiomar, Phys. Rev. D \textbf{90,} 082002\textbf{\ } (2014). [arXiv:1409.2022]

\bibitem {k15}A. \"{O}vg\"{u}n, K. Jusufi and I. Sakall{\i}, Phys. Rev. D \textbf{99}, 024042 (2019). [arXiv:1804.09911]

\bibitem {l15}R. Bluhm, N.L. Gagne, R. Potting and A. Vrublevskis, Phys. Rev.
D \textbf{77}, 125007 (2008).

\bibitem {l14}R. Casana, A. Cavalcante, F. P. Poulis, E. B. Santos Phys. Rev. D \textbf{97}, 104001 (2018).

\bibitem {d14}R. J. Yang, H. Gao, Y. G. Zheng, W. Qin, Commun.Theor. Phys.
\textbf{71}, 568 (2019).

\bibitem {16}A. \"{O}vg\"{u}n, K. Jusufi and I. Sakalli, Annals Phys. 399,\textbf{193} (2018).

\bibitem {17}S. Kanzi and I. Sakalli, Nucl. Phys. B \textbf{946}, 114703 (2019).  [arXiv:1905.00477]

\bibitem {18}C. Ding, C. Liu, R. Casana and A. Cavalcante, Eur. Phys. J. C
\textbf{80}, 178 (2020). [arXiv:1910.02674]

\bibitem {19}S.W. Hawking, Nature \textbf{248}, 30 (1974).
https://doi.org/10.1038/ 248030a0

\bibitem {20}J. Maldacena, A. Strominger. Phys. Rev. D \textbf{55}, 861 (1997).

\bibitem {H21}S. W. Hawking, Nature \textbf{248}, 30 (1974).

\bibitem {H22}S. W. Hawking, Commun. Math. Phys. \textbf{43}, 199 (1975).

\bibitem {21}D. N. Page, Phys. Rev. D \textbf{13}, 198 (1976).

\bibitem {22}D. N. Page, Phys. Rev. D\textbf{14}, 3260 (1976).

\bibitem {23}K. Jusufi, M. Amir ,M. Sabir Ali ,S. D. Maharaj, Phys. Rev. D
\textbf{102}, 064020 (2020).

\bibitem {24}M. K. Parikh, F. Wilczek, Phys. Rev. Lett. \textbf{85}, 5049 (2000).

\bibitem {25}S. Fernando, Gen. Relativ. Gravit. \textbf{37}, 461 (2005).

\bibitem {26}W. Kim and J.J. Oh, JKPS \textbf{52}, 986 (2008).

\bibitem {27}T. Harmark, J. Natario, R. Schiappa, Adv. Theor. Math. Phys.
\textbf{14}, 727 (2010).

\bibitem {28}P. Boonserm, T. Ngampitipan, P. Wongjun, Eur. Phys. J. C
\textbf{79}, 330 (2019). https://doi.org/10.1140

\bibitem {y28}I. Sakalli, Phys. Rev. D \textbf{94}, 084040 (2016).
[arXiv:1606.00896 [gr-qc]]

\bibitem {d28}A. Al-Badawi, I. Sakalli, and S. Kanzi, Annals Phys.
\textbf{412}, 168026 (2020).

\bibitem {h28}A. Al-Badawi, S. Kanzi, I. Sakalli, Eur. Phys. J. Plus
\textbf{135},(2020). 

\bibitem {g28}H. Gursel and I. Sakalli, Eur. Phys. J. C \textbf{80}, 234 (2020).

\bibitem {l28}S. Kanzi, S. H. Mazharimousavi, I. Sakalli, Annals Phys. (2020).

\bibitem {s28}R. Bluhm and V. A. Kosteleck%
\'{}%
y, Phys. Rev. D \textbf{71}, 065008 (2005).

\bibitem {d29}R. A. Konoplya, A. Zhidenko, Rev. Mod. Phys.\textbf{83}, 793 (2011).

\bibitem {u29}R. A. Konoplya, Phys. Rev. D \textbf{68}, 024018 (2003).

\bibitem {f29}S. L. Detweiler, Astrophys. J. \textbf{239}, 292 (1980).

\bibitem {s29}E. Berti, V. Cardoso, and C. M. Will, Phys. Rev. D \textbf{73},
064030 (2006).

\bibitem {Sara34}S. Iyer, Phys. Rev. D \textbf{35}, 3632 (1987).

\bibitem {Sara35}S. Iyer, C. M. Will, Phys. Rev. D \textbf{35}, 3621 (1987).

\bibitem {r29}G.T. Horowitz and V. Hubeny, Phys. Rev. D \textbf{62}, 024027 (2000).

\bibitem {t29}D. Birmingham, I. Sachs, and S.N. Solodukhin, Phys. Rev.Lett.
\textbf{88}, 151301 (2002).

\bibitem {tr29}V. Cardoso and J.P.S. Lemos, Phys. Rev. D \textbf{63}, 124015 (2001).

\bibitem {g29}L. Huang, J. Chen,Y. Wang, Eur. Phys. J. C \textbf{78}, 229 (2018).

\bibitem {w29}G. Kunstatter, Phys. Rev. Lett. \textbf{90}, 161301 (2003).

\bibitem {x29}S. Hod, Phys. Rev. D \textbf{59}, 024014 (1999).

\bibitem {z29}Y. Hatsuda, M. Kimura, Pheys. Rev. D \textbf{102}, 044032 (2020).

\bibitem {j29}G. B. Cook, M. Zalutskiy, Pheys. Rev. D \textbf{90}, 124021 (2014).

\bibitem {b29}P. Pani, E. Berti, L. Gualtieri, Phys. Rev. Lett. \textbf{110},
241103 (2013).

\bibitem {29}S. Chandrasekhar, The Mathematical Theory of Black Holes, Oxford
University Press, Oxford, UK, (1983).

\bibitem {Sara30}Handbook of Mathematical functions, edited by M. Abramowitz
and I. A. Stegun (Dover, New York,1970).

\bibitem {Sara31}S. Detweiler, Phys. Rev. D \textbf{22}, (1980).

\bibitem {Sara32}A. A. Starobinskii, Zh. Eksp. Teor. Fiz \textbf{64}, 48
(1973). [Sov. Phys. JETP 37, 28 (1973)].

\bibitem {Sara39}E. Newman, R. Penrose, J. Math. Phys. \textbf{3}, 566 (1962 ).

\bibitem {Sara37}B. Chen, L. C. Stein, Phys. Rev. D \textbf{96}, 064017 (2017).

\bibitem {Sara38}B. D. Nikoli\'{c}, M. R. Panti\'{c}, [arXiv:1210.5922], (2013).

\bibitem {spinw}G.F. Torres del Castillo, $3-D$ Spinors, Spin-Weighted Functions and their Applications, (Birkhäuser, Boston, 2003).

\bibitem {Sara33}Y.G. Miao, Z.M. Xu, Phys. Lett. B \textbf{772}, 542 (2017).

\bibitem {asym}J. López, J. Comput. Appl. Math. \textbf{102}, 181 (1999).

\bibitem {Sarak34}B. F. Schutz, C. M. Will, Astrophys. J. Lett.  \textbf{291},  L33-L36 (1985).

\bibitem {Sarak35}E. Seidel, S. Iyer, Phys. Rev. D.  \textbf{41},374 (1990).

\bibitem {Sarak36}W. A. Carlson, A. s. Cornell, B. Jordan, [arXiv:1201.3267 ], (2012). 

\bibitem {data} A.F. Zinhailo, Eur. Phys. J. C \textbf{78}, 992 (2018).

\bibitem{wkb1}R.A. Konoplya, A. Zhidenko, and A.F. Zinhailo, Class. Quantum Grav., \textbf{36}, 155002 (2019).

\end{thebibliography}
\end{document}